\documentclass[12pt,a4paper]{article}
\usepackage{amsmath,amssymb,amsthm}
\usepackage{graphicx}
\usepackage{cite}

\renewcommand{\baselinestretch}{1.5}

\begin{document}

\title{{\Large \textbf{Magneto-optical Selection Rules in Bilayer Bernal
Graphene}}\\
}
\author{ Yen-Hung Ho$^{1,2}$, Yu-Huang{\ Chiu$^{1,}\thanks{%
Corresponding author: \textit{E-mail address: airegg.py90g@nctu.edu.tw}}$}~,
De-Hone Lin$^{2}$, Chen-Peng Chang$^{3}$, \and and Ming-Fa Lin$^{1,}\thanks{%
Corresponding author: \textit{E-mail address: mflin@mail.ncku.edu.tw}}$ \\
%EndAName
{\footnotesize $^{1}$Department of Physics, National Cheng Kung University,
Tainan, Taiwan 701}\\
{\footnotesize $^{2}$Department of Physics, National Sun Yat-Sen University,
Kaohsiung, Taiwan 804}\\
{\footnotesize $^{3}$Center of General Education, National Tainan University
of Technology, Tainan, Taiwan 710}\\
}
\date{\today}
\maketitle

\begin{abstract}
The low-frequency magneto-optical properties of bilayer Bernal graphene are
studied by the tight-binding model with four most important interlayer
interactions taken into account. Since the main features of the wave
functions are well depicted, the Landau levels can be divided into two
groups based on the characteristics of the wave functions. These Landau
levels lead to four categories of absorption peaks in the optical absorption
spectra. Such absorption peaks own complex optical selection rules and these
rules can be reasonably explained by the characteristics of the wave
functions. In addition, twin-peak structures, regular frequency-dependent
absorption rates and complex field-dependent frequencies are also obtained
in this work. The main features of the absorption peaks are very different
from those in monolayer graphene and have their origin in the interlayer
interactions. \vskip0.5 truecm

 \vskip0.2 truecm
%\noindent \textbf{\textrm{PACS numbers:}} {78.20.Ls, 78.67.Pt, 73.20.At, 71.15.Dx} \pagebreak
DOI: 10.1021/nn9015339, ACSNano \textbf{4}, 1465 (2010)
\pagebreak
\end{abstract}

\renewcommand{\baselinestretch}{1.4}

\newpage \renewcommand{\baselinestretch}{1.5}

Ever since few-layer graphenes (FLGs) have first been produced by mechanical
exfoliation$^{1,2}$ and thermal decomposition,$^{3,4}$ they have attracted
considerable experimental and theoretical research because of their exotic
physical properties resulting from the hexagonal structure. Monolayer
graphene (MG) and bilayer Bernal graphene (BBG; AB-stacked bilayer
graphene), two of the simplest few-layer graphenes, display very different
electronic structures. MG exhibits isotropic linear bands near the Fermi
level $E_{F}=0$; these bands become gradually anisotropic parabolic bands as
the energy exceeds the region of $\pm $0.5 eV.$^{5}$ In BBG, the linear
bands change into two pairs of parabolic bands due to the present interlayer
interactions and stacking configurations.$^{6-14}$ Each pair is comprised of
the occupied valence and unoccupied conduction bands, which are asymmetric
about $E_{F}$.$^{6,9}$ The conduction and valence bands of the first pair
slightly overlap about $E_{F}$, while those of the second pair rise at the
energies$^{12}$ 0.34 eV ($\simeq \gamma _{1}+\gamma _{6}$) and -0.39 eV ($%
\simeq -\gamma _{1}+\gamma _{6}$), respectively, where $\gamma _{s}$'s$%
^{8,10,15-17}$ ($s$=1, 3, 4 and 6) are four important interlayer
interactions in BBG. The unusual electronic structures of MG and BBG have
been experimentally verified through angular-resolved photoemission
spectroscopy.$^{18,19}$ The main features of the zero-field energy bands,
\textit{e.g.}, the broken linear dispersions in MG and the two groups of
asymmetric electronic states in BBG, should be reflected in the features of
the Landau levels (LLs) resulting from the magnetic fields.$^{12}$

In the presence of a uniform perpendicular magnetic field $\mathbf{B}=B_{0}%
\hat{z}$, the zero-field energy bands of FLGs become the dispersionless LLs.
Some magneto-electronic properties of FLGs have been obtained by
measurements of the quantum Hall conductivity,$^{2,20-22}$ scanning
tunneling spectroscopy,$^{22-24}$ and cyclotron resonance.$^{25,26}$ The
effective-mass approximation (EM) and the tight-binding model (TB), two
theoretical methods based on different concepts in calculations, are usually
utilized to study the magneto-optical properties of layered-graphene
systems. In MG, EM shows that the linear bands become LLs characterized by
the special relation $E_{n}^{c,v}\propto \sqrt{n^{c,v}B_{0}}$,$^{27,28}$
where $c$ ($v$) is the index of the conduction (valence) bands. $E_{n}^{c.v}$
and $n^{c,v}$ are the energy and the integer quantum number of the $n$th
energy band respectively. However, TB indicates that the special relation
will be broken as the energy exceeds the region of $\pm $0.5 eV since the
linear dispersions become gradually parabolic as energy exceeds this region.$%
^{29}$ As for the optical-absorption spectra, the low-frequency absorption
peaks possess the selection rule $\left\vert \triangle n\right\vert
=\left\vert n^{c}-n^{v}\right\vert =1$ and their frequency is proportional
to $\sqrt{B_{0}}$,$^{30}$ which has been confirmed by magneto-transmission
measurements.$^{25,31,32}$

For BBG, EM takes the LL structures of MG and uses them in the BBG
calculations. Interlayer interactions are regarded as perturbations (EM
considers only one or two interlayer interactions). Based on EM
calculations, the first theoretical study$^{13}$ of LLs in BBG derived an
effective two-dimensional Hamiltonian that describes the low-energy
electronic excitations. Furthermore, EM predicts the existence of two groups
of the occupied and unoccupied LLs, and they are symmetric about $E_{F}$
because the parabolic bands at $B_{0}$=0\ are symmetric in the calculations.$%
^{30,33,34}$ The low-energy LLs in their results follow the relation $%
E_{n}^{c,v}\propto B_{0}$.$^{30}$ As for the optical absorption spectra, EM
calculations show that there are three categories of absorption peaks.$^{30}$
The absorption frequencies obey a linear function of $B_{0}$ in weak fields,
and depend linearly on $\sqrt{B_{0}}$ as the corresponding energy leaves the
parabolic band region.$^{30}$ The three categories own the same optical
selection rule $\left\vert \triangle n\right\vert =1$.$^{14,30,35,36}$ In
TB, a magnetic field and four most important interlayer interactions$%
^{8,12,13}$ are simultaneously included in the calculations. The magnetic
flux $\phi $\ through a hexagon corresponding to the strongest field
strength ($\sim $60 T) in experiments is much smaller than the flux quantum $%
\phi _{0}$ ($\phi \sim \phi _{0}$/1320). From the viewpoint of the TB
perturbation theory, electronic states that lie in close proximity to each
other at $B_{0}$=0 will be aggregated and become dispersionless LLs under
the influence of a magnetic field. The characteristics of electronic states
at $B_{0}$=0 are directly reflected in the main features of LLs. The
interlayer interactions strongly affect the electronic properties,
especially for low-energy levels. These interactions are neither negligible
nor suitable as perturbations. TB indicates that the low-energy LLs should
exhibit more complex features;$^{12}$ \textit{i.e.}, there exists an energy
gap, two groups of asymmetric LLs ($^{1st}$LL$^{c,v}$s and $^{2nd}$LL$^{c,v}$%
s), and no simple relations between $E_{n}^{c,v}$ and $B_{0}$. The main
features would be reflected in the magneto-optical properties.

In the initial stage of our works in studying the magneto-electronic
properties, only eigenvalues and eigenvectors at strong magnetic field
strength were obtained$^{37}$ because the Hamiltonian matrix is quite large,
\textit{e.g.}, the matrix is about 31600$\times $31600 for MG at 10 T.
Through the band-like-matrix strategy$^{12,38-42}$ designed by our group,
the eigenvalues and the wave functions at weak fields ($\sim $1 T) are thus
clearly depicted.$^{12}$ However, the calculation of optical properties is
more complex and time-consuming than that of the electronic properties since
detailed calculations for a huge velocity matrix and reliable
characterization of the wave functions are necessary for the TB-based
optical-absorption spectra. In this work, the localized features of the wave
functions are utilized to effectively reduce the computation time, and thus
the optical absorption spectra can be solved. It's the first time that the
optical properties of BBG calculated by TB are obtained. The influence of
the four most important interlayer interactions on the magneto-optical
absorption spectra of BBG is investigated in detail by TB and gradient
approximation.$^{37,40-44}$ The wave functions are clearly depicted and used
to divide the LLs into two groups,$^{12}$ $^{1st}$LLs and $^{2nd}$LLs. The
two groups lead to four categories of optical absorption peaks. These peaks
exhibit twin-peak structures, complex selection rules, regular
frequency-dependent absorption rates, and complicated field-dependent
absorption frequencies. What deserves to be mentioned is that our study is
more complete in nature and the presented results are more reliable than
those of previous works. Previously experimental and theoretical research
mainly focused on the optical properties associated with $^{1st}$LLs.
However, in our work, the optical absorption peaks resulting from both
groups are discussed in detail and each peak can be clearly identified. A
detailed comparison of optical-absorption spectra between MG and BBG is
made. In addition, comparisons of optical properties between AA-stacked
bilayer graphene (AABG) and BBG and between EM and TB are also made.

\begin{flushleft}
\textbf{RESULTS AND DISCUSSION}
\end{flushleft}

The dispersionless LLs of BBG at $B_{0}=40$ T are shown in Figure 1a. The
interlayer interactions strongly influence the main features of LLs. The
conduction and valence bands are asymmetric. There is an energy gap between
the lowest unoccupied LL and the highest occupied LL. Based on the
characteristics of wave functions, LLs can be classified into two groups, $%
^{1st}$LLs (black curves) and $^{2nd}$LLs (red curves). $^{1st}$LLs appears
at $\left\vert E^{c,v}\right\vert >0$, and $^{2nd}$LLs begins at $%
E^{c}\gtrsim \gamma _{1}+\gamma _{6}$ (0.34 eV) and $E^{v}\lesssim -\gamma
_{1}+\gamma _{6}$ (-0.39 eV). The onset energies of the first and second
groups are consistent with those of the first and second pairs in the
zero-field subbands. Furthermore, no simple relation between the energy and
field strength can be found for these LLs, \textit{i.e.}, $E_{n}^{c,v}$ is
not directly proportional to either $B_{0}$ or $\sqrt{B_{0}}$.$^{12}$

The characteristics of the wave functions deserve a closer investigation
because they strongly dominate the optical properties. Each LL is composed
of fourfold degenerate states with similar characteristics. At $%
k_{x}=k_{y}=0 $, the wave functions mainly consist of four localized
subenvelope functions centered at positions $x_{1}=1/6$, $x_{2}=1/3$, $%
x_{3}=2/3$, and $x_{4}=5/6$. $x_{j=1-4}=A_{i,N}/2R_{B}$ or $=B_{i,N}/2R_{B}$%
. $A_{i,N}$ ($B_{i,N}$) indicates the $N$th $A$ ($B$) atom on the $i$th
layer, where $N=1,2...2R_{B}$ and $i=1,2$ denote the atomic site in the
primitive unit cell and the layer index respectively (please see the
detailed discussion regarding these representations in the METHODS section).
The feature of a wave function resulting from the atoms with an even number (%
$A_{i,e}^{c,v}$ and $B_{i,e}^{c,v}$) is similar to that with an odd ($%
A_{i,o}^{c,v}$ and $B_{i,o}^{c,v}$) number, where $e$ ($o$) is an even (odd)
integer. Because of this, only odd atoms are discussed in the following
calculations. For convenience, one of the fourfold degenerate states at $%
x_{1}=1/6$ is selected to be examined, as shown in Figures 1b-1e. Through
appropriate fitting, the wave functions of the $n_{1}$th $^{1st}$LL (black
curves) and $n_{2}$th $^{2nd}$LL (red curves) can be expressed as
\begin{subequations}
\begin{align}
A_{1,o}^{c}\text{, }A_{2,o}^{c}& \propto \varphi _{2}\left( x\right) \text{,
}B_{1,o}^{c}\propto \varphi _{1}\left( x\right) \text{, }B_{2,o}^{c}\propto
\varphi _{0}\left( x\right) \text{, }n_{1}^{c,eff}=0\text{ for }n_{1}=1\text{%
;} \\
A_{1,o}^{v}\text{, }A_{2,o}^{v}& \propto \varphi _{0}\left( x\right) \text{,
}B_{1,o}^{v}\propto \varphi _{2}\left( x\right) \text{, }B_{2,o}^{v}\propto
\varphi _{1}\left( x\right) \text{, }n_{1}^{v,eff}=1\text{ for }n_{1}=1\text{%
;} \\
A_{1,o}^{c,v}\text{, }A_{2,o}^{c,v}& \propto \varphi _{n_{1}-1}\left(
x\right) \text{, }B_{1,o}^{c,v}\propto \varphi _{n_{1}-2}\left( x\right)
\text{, }B_{2,o}^{c,v}\propto \varphi _{n_{1}}\left( x\right) \text{ for }%
n_{1}^{c,v,eff}=n_{1}\geq 2\text{.}
\end{align}%
\end{subequations}
\begin{subequations}
\begin{eqnarray}
A_{1,o}^{c,v}\text{, }A_{2,o}^{c,v} &\propto &\varphi _{0}\left( x\right)
\text{, }B_{1,o}^{c,v}\propto \varphi _{2}\left( x\right) \text{, }%
B_{2,o}^{c,v}\propto \varphi _{1}\left( x\right) \text{, }n_{2}^{c,v,eff}=0%
\text{ for }n_{2}=1\text{;} \\
A_{1,o}^{c,v}\text{, }A_{2,o}^{c,v} &\propto &\varphi _{n_{2}-1}\left(
x\right) \text{, }B_{1,o}^{c,v}\propto \varphi _{n_{2}-2}\left( x\right)
\text{, }B_{2,o}^{c,v}\propto \varphi _{n_{2}}\left( x\right) \text{ for }%
n_{2}^{c,v,eff}=n_{2}-1\geq 1\text{.}
\end{eqnarray}%
$\varphi _{n}\left( x\right) $ is the product of the $n$th-order Hermite
polynomial and Gaussian function,$^{12,29,40}$ where $n$ is the number of
zero points of $\varphi _{n}\left( x\right) $ and chosen to define the
quantum number of a LL.$^{12}$ That is to say, there are four quantum
numbers in each LL since a LL wave function is made up of four magnetic TB
functions. Thus an effective quantum number, $n^{eff}$, defined by one of
these four quantum numbers is necessary to identify each LL. The effective
quantum number of a LL in each group is assigned by the number of zero
points in one of the four magnetic TB functions, with the largest amplitude
among these four TB functions at the onset energy of each group.
Accordingly, $n_{1}^{c,v,eff}$ ($n_{2}^{c,v,eff}$) is the effective quantum
number of the $^{1st}$LLs ($^{2nd}$LLs) defined by the quantum number of $%
B_{2,o}^{c,v}$ ( $A_{1,o}^{c,v}$). It should be noted that the wave
functions of two groups are composed of not only the $B$ atoms but also the $%
A$ atoms, which are different from those calculated by EM including only
contributions from the $B$ atoms.$^{13,14}$ Such difference could cause the
optical properties obtained by TB and EM to exhibit different features. $%
^{1st}$LLs and $^{2nd}$LLs can be distinguished by the different features of
the wave functions in the mixed region ($E^{c}\gtrsim 0.34$ eV and $%
E^{v}\lesssim -0.39$ eV), which is useful in analyzing the optical
excitation channels.

When BBG is subjected to an electromagnetic field at zero temperature, only
excitations from the occupied to the unoccupied bands occur. Based on
Fermi's golden rule, the optical-absorption function is given by
\end{subequations}
\begin{eqnarray}
A(\omega ) &\propto &\sum_{n_{1,2}^{c,eff},n_{1,2}^{v,eff}}\int_{1stBZ}{%
\frac{d\mathbf{k}}{(2\pi )^{2}}}|\langle \psi ^{c}(n_{1,2}^{c,eff},\mathbf{k}%
)|{\frac{\widehat{\mathbf{E}}\cdot \mathbf{P}}{m_{e}}}|\psi
^{v}(n_{1,2}^{v,eff},\mathbf{k})\rangle |^{2}  \notag \\
&&\times Im\{\frac{f\left[ E^{c}\left( \mathbf{k},n_{1,2}^{c,eff}\right) %
\right] -f\left[ E^{v}\left( \mathbf{k},n_{1,2}^{v,eff}\right) \right] }{%
E^{c}\left( \mathbf{k},n_{1,2}^{c,eff}\right) -E^{v}\left( \mathbf{k}%
,n_{1,2}^{v,eff}\right) -\omega -i\Gamma }\}\text{,}
\end{eqnarray}%
where $\widehat{\mathbf{E}}$ is the unit vector of an electric polarization.
$\Gamma $ (=1 meV) is a broadening parameter and often affected by
temperature and defect effects. If a purer sample is manufactured and
observed its physical properties under a sufficiently low temperature, $%
\Gamma $ will be small enough for observing important fine structures.$%
^{22,45}$ Furthermore, $\langle \psi ^{c}(n_{1,2}^{c},\mathbf{k})|\widehat{%
\mathbf{E}}\cdot \mathbf{P/}m_{e}|\psi ^{v}(n_{1,2}^{v},\mathbf{k})\rangle $
is the velocity matrix element derived from the dipole transition and
denoted $M^{cv}$. In this work, $M^{cv}$ is calculated by gradient
approximation. Through detailed calculations, $M^{cv}$ can be simplified to
be the product of three matrices, the initial state (occupied state; $\psi
^{v}$), the final state (unoccupied state; $\psi ^{c}$) and $\nabla
_{k}H_{ij}$. The last term corresponds to the direction of electric
polarization, \textit{i.e.}, it is $\partial H_{ij}/\partial k_{x}$ for
polarization along the x-axis in this work. The elements of $\partial
H_{ij}/\partial k_{x}$ are non-zero only for $H_{ij}$ associated with the
hopping integrals, as are the velocity matrix elements. In other words, when
the velocity matrix element does not vanish, the initial and final states in
the product should be the two states corresponding to the non-vanishing
hopping integrals. Accordingly, $M^{cv}$ is expressed as
\begin{eqnarray}
&&\{\sum_{i=1,2}\sum_{N,N^{\prime }=1}^{2R_{B}}[\left(
A_{io}^{c}+A_{ie}^{c}\right) ^{\ast }\times \left( B_{io^{\prime
}}^{v}+B_{ie^{\prime }}^{v}\right) ]\nabla _{k}\left\langle A_{i,N\mathbf{k}%
}\left\vert H_{B}\right\vert B_{i,N^{\prime }\mathbf{k}}\right\rangle  \notag
\\
&&+\sum_{i,j=1,2;i\neq j}\sum_{N,N^{\prime }=1}^{2R_{B}}[\left(
B_{io}^{c}+B_{ie}^{c}\right) ^{\ast }\times \left( B_{jo^{\prime
}}^{v}+B_{je^{\prime }}^{v}\right) ]\nabla _{k}\left\langle B_{i,N\mathbf{k}%
}\left\vert H_{B}\right\vert B_{j,N^{\prime }\mathbf{k}}\right\rangle  \notag
\\
&&+\sum_{i,j=1,2;i\neq j}\sum_{N,N^{\prime }=1}^{2R_{B}}[\left(
B_{io}^{c}+B_{ie}^{c}\right) ^{\ast }\times \left( A_{jo^{\prime
}}^{v}+A_{je^{\prime }}^{v}\right) ]\nabla _{k}\left\langle B_{i,N\mathbf{k}%
}\left\vert H_{B}\right\vert A_{j,N^{\prime }\mathbf{k}}\right\rangle \}+h.c.
\end{eqnarray}%
The first, second, and third term in eq 4 correspond to the three hopping
integrals $\gamma _{0}$, $\gamma _{3}$, and $\gamma _{4}$, respectively.
That is to say, the optical excitations are provided by the three channels
related to $\gamma _{0}$, $\gamma _{3}$ and $\gamma _{4}$. For convenience, $%
\left( A_{io}^{c}+A_{ie}^{c}\right) ^{\ast }\times \left( B_{io^{\prime
}}^{v}+B_{ie^{\prime }}^{v}\right) $, $\left( B_{io}^{c}+B_{ie}^{c}\right)
^{\ast }\times \left( B_{jo^{\prime }}^{v}+B_{je^{\prime }}^{v}\right) $,
and $\left( B_{io}^{c}+B_{ie}^{c}\right) ^{\ast }\times \left( A_{jo^{\prime
}}^{v}+A_{je^{\prime }}^{v}\right) $\ are represented by $M_{AB}^{cv}(\gamma
_{0})$, $M_{BB}^{cv}(\gamma _{3})$, and $M_{BA}^{cv}(\gamma _{4})$,
respectively in the following discussions. It should be noted that eq 4
implies that the characteristics of the wave functions would play a dominant
role in determining the selection rule and absorption rate in optical
excitations.

The low-frequency optical-absorption spectra of BBG regarding four different
magnetic field strength are shown in Figures 2a-2d, respectively. The
absorption spectrum of MG is also shown in Figure 2e. In Figure 2a, the
absorption spectrum of BBG at $B_{0}=40$ T presents prominent and
inconspicuous (indicated by the black arrows) peaks. The prominent peaks
with definite optical selection rules are discussed in detail in the
following paragraphs. The peaks can mainly be classified into four
categories of peaks, $\omega _{11}$'s (black dots), $\omega _{22}$'s (red
dots), $\omega _{12}$'s (green dots), and $\omega _{21}$'s (blue dots),
which originate in the four different excitation channels, $^{1st}$LL$^{v}$s
to $^{1st}$LL$^{c}$s, $^{2nd}$LL$^{v}$s to $^{2nd}$LL$^{c}$s, $^{1st}$LL$%
^{v} $s to $^{2nd}$LL$^{c}$s, and $^{2nd}$LL$^{v}$s to $^{1st}$LL$^{c}$s
respectively. The former (latter) two channels result from the transitions
between two LLs in the same (different) groups, as illustrated by the inset
of Figure 2d, and they display obvious optical selection rules.

The optical excitations associated with each prominent peak can clearly be
identified, as shown in Figure 2a. For convenience, the excitations of $%
\omega _{11}$'s, $\omega _{22}$'s, $\omega _{12}$'s, and $\omega _{21}$'s
are respectively represented as $n_{1}^{v,eff}\rightarrow n_{1}^{c,eff}$, $%
n_{2}^{v,eff}\rightarrow n_{2}^{c,eff}$, $n_{1}^{v,eff}\rightarrow
n_{2}^{c,eff}$, and $n_{2}^{v,eff}\rightarrow n_{1}^{c,eff}$ in the
following. The selection rule of $\omega _{ij}$'s is denoted by $\triangle
n_{ij}$ ($=n_{j}^{c,eff}-n_{i}^{v,eff}$). As to $\omega _{11}$'s and $\omega
_{22}$'s, only the first peak ($\omega _{11}^{1}$) of $\omega _{11}$'s
originates in single channel $1\rightarrow 2$. The other $m$th peak $\omega
_{11}^{m}$ ($\omega _{22}^{m}$) consists of the pair $\omega _{11}^{m,L}$
and $\omega _{11}^{m,H}$ ($\omega _{22}^{m,L}$ and $\omega _{22}^{m,H}$)
corresponding to $m\rightarrow m+1$ and $m+1\rightarrow m$ ($m-1\rightarrow m
$ and $m\rightarrow m-1$) respectively. The superscript $L$ ($H$) indicates
the lower (higher) energy of the pair. For example, $\omega _{11}^{2}$ ($%
\omega _{22}^{1}$) is composed of $\omega _{11}^{2,L}$ and $\omega
_{11}^{2,H}$ ($\omega _{22}^{1,L}$ and $\omega _{22}^{1,H}$) owing to $%
2\rightarrow 3$ and $3\rightarrow 2$ ($0\rightarrow 1$ and $1\rightarrow 0$)
respectively. The origin of a twin-peak structure is that the energies of $%
m\rightarrow m+1$ ($m-1\rightarrow m$) and $m+1\rightarrow m$ ($m\rightarrow
m-1$) in $\omega _{11}$'s ($\omega _{22}$'s) are slightly different owing to
the asymmetry of the occupied and unoccupied LLs. It should be noted that
similar twin-peak structures are also shown in a recent experiment conducted
to study the magneto-absorption spectra.$^{45}$ In short, the optical
selection rules of $\omega _{11}$'s and $\omega _{22}$'s can be represented
by $\triangle n_{11}=\triangle n_{22}=\pm 1$, which is the same as the LLs
in MG.

As for the intergroup transitions, twin-peak structures are also obtained in
$\omega _{12}$'s and $\omega _{21}$'s. The $m$th peak of $\omega _{12}$'s is
formed with the pair, $\omega _{12}^{m,L}$ originating in $m\rightarrow m$
and $\omega _{12}^{m,H}$ resulting from $m+1\rightarrow m-1$. For instance,
the first pair is indicated by $\omega _{12}^{1,L}$ and $\omega _{12}^{1,H}$
in Figure 2a, which correspond to $1\rightarrow 1$ and $2\rightarrow 0$
respectively. That is to say, $\omega _{12}^{m,L}$ owns the optical
selection rule $\triangle n_{12}=0$ and $\omega _{12}^{m,H}$ possesses $%
\triangle n_{12}=-2$. For $\omega _{21}$'s, the excitations $0\rightarrow 0$
and $0\rightarrow 2$ lead to the peaks $\omega _{21}^{1,L}$ and $\omega
_{21}^{1,H}$ of the first pair in Figure 2a respectively. The other channels
$m-1\rightarrow m+1$ and $m\rightarrow m$ result in the pair $\omega
_{21}^{m,L}$ and $\omega _{21}^{m,H}$ respectively. In other words, the
selection rules of $\omega _{21}$'s are $\triangle n_{21}=0$ and $\triangle
n_{21}=2$. Obviously, the optical selection rules of $\omega _{12}$'s and $%
\omega _{21}$'s are different from those of $\omega _{11}$'s and $\omega
_{22}$'s, a fact which is not clearly discussed in the previous theoretical
works.

The optical selection rules can be explained from two different aspects,
\textit{i.e.}, the effective quantum numbers and the numbers of zero points
between the initial and final states. Both aspects are based on the premise
that the velocity matrix is derived from the dipole transition. The product
of the conduction and valence wave functions in eq 4 would not only
determine the allowed excitation channel, but also enable us to estimate the
absorption rate. This means that the dissimilarities in selection rules of
the four categories mainly result from the characteristics of the wave
functions. In eq 4, $M_{AB}^{cv}(\gamma _{0})$ associated with $\gamma _{0}$
dominates the excitations of the prominent peaks. It strongly depends on the
numbers of zero points of $A_{1}^{c,v}$ and $B_{1}^{c,v}$ ($A_{2}^{c,v}$ and
$B_{2}^{c,v}$) since $M_{AB}^{cv}(\gamma _{0})=$ $\left(
A_{io}^{c}+A_{ie}^{c}\right) ^{\ast }\times \left( B_{io^{\prime
}}^{v}+B_{ie^{\prime }}^{v}\right) $. $M_{AB}^{cv}(\gamma _{0})$ has
non-zero values only when $A_{1}^{c,v}$ and $B_{1}^{c,v}$ ($A_{2}^{c,v}$ and
$B_{2}^{c,v}$) own the same $\varphi _{n}(x)$ because of the orthogonality
of $\varphi _{n}(x)$. Thus, the selection rules of prominent peaks can be
easily obtained by the definition of the wave functions in eqs 1 and 2. From
the aspect of the numbers of zero points between $A_{i}^{c,v}$ and $%
B_{i}^{c,v}$, these states own the same number of zero points in an
available excitation channel in both intragroup and intergroup transitions.
As for the effective quantum numbers, the selection rules are $\triangle
n_{11}=\triangle n_{22}=\pm 1$, $\triangle n_{12}=0$, $-2$, and $\triangle
n_{21}=0$, $2$. Although these selection rules are different, all
excitations obey the underlying condition that the numbers of zero points of
the initial and final states in the dipole-transition-derived velocity
matrix are the same.

In addition to the optical selection rules, the peak intensity and
absorption frequency are also discussed in this work. The absorption rate
significantly relies on the field strength and the excitation energy. In
Figures 2a-2d, the peak intensity is increased with enlarging field
strength, while the opposite is true for the peak number. The peak intensity
under a fixed magnetic field strength is determined by the product of two
wave functions with respect to the two nearest-neighbor atoms. The wave
functions of $^{1st}$LLs ($^{2nd}$LLs) are mainly dominated by the $B$ ($A$)
atoms. The amplitudes of $A_{i}^{c,v}$ and $B_{i}^{c,v}$, respectively,
gradually grow and decrease (decrease and grow) with increasing energy in $%
^{1st}$LLs ($^{2nd}$LLs). Accordingly, the value of the product of $%
A_{i}^{c,v}$ and $B_{i}^{c,v}$ increases (decreases) with rising energy for $%
\omega _{11}$'s and $\omega _{22}$'s ($\omega _{12}$'s and $\omega _{21}$%
's). This value is largest when the amplitudes of $A_{i}^{c,v}$ and $%
B_{i}^{c,v}$ are equal. The intensities of $\omega _{11}$'s and $\omega
_{22} $'s ($\omega _{12}$'s and $\omega _{21}$'s) thus increase (decline)
with increasing frequency. Furthermore, some peaks with commensurate
energies overlap because the peak spacing declines as the frequency
increases; this overlap is responsible for the variation in peak intensity.
In contrast to BBG, MG exhibits absorption peaks with uniform intensity
(Figure 2e).

The field-dependent absorption frequencies ($\omega _{a}$'s) of $\omega
_{11} $'s, $\omega _{22}$'s, $\omega _{12}$'s and $\omega _{21}$'s are shown
in Figure 3a by the black, red, green, and blue dots, respectively. $\omega
_{a} $'s of the four categories rise with increasing $B_{0}$. These
absorption frequencies do not exhibit the simple relations indicated by EM
calculations,$^{29}$ \textit{i.e.}, the frequencies are not directly
proportional to either $B_{0}$ or $\sqrt{B_{0}}$. Apparently, $\omega
_{a}\propto \sqrt{B_{0}}$ in MG (Figure 3b) is very different from that in
BBG. The main reason for the difference between BBG and MG is that
interlayer interactions play important roles in BBG and are not negligible
in calculations. In addition, the convergent frequencies of $\omega _{11}$%
's, $\omega _{22}$'s, $\omega _{12}$'s and $\omega _{21}$'s at the weak
field strength are approximately 0, 0.73 eV (2$\gamma _{1}$), 0.34 eV ($%
\gamma _{1}+\gamma _{6}$), and 0.39 eV ($\gamma _{1}-\gamma _{6}$)
respectively. This implies that the optical measurements can reasonably
determine the values of $\gamma _{1}$ and $\gamma _{6}$ through observing
the convergent frequencies of absorption peaks. The predicted results are
very useful and reliable for experimental studies.

The inconspicuous peaks indicated by arrows in Figure 2a also obey specific
selection rules (not shown). These peaks mainly originate from excitations
related to two interlayer interactions ($\gamma _{3}$ and $\gamma _{4}$) and
differ from the prominent peaks associated with the intralayer hopping
integral ($\gamma _{0}$). The inconspicuous peak intensities are much weaker
than the intensities of the prominent peaks since the values of the velocity
matrix elements corresponding to $\gamma _{3}$ and $\gamma _{4}$ are much
smaller than those related to $\gamma _{0}$. The different origins and
selection rules of the prominent and inconspicuous peaks imply that the
optical properties can not be simply derived from the eigenenergies of LLs.
The detailed calculations of the velocity matrix and reliable
characterization of the wave functions are necessary to determine the
optical selection rules and the peak intensity.

The atomic hopping integrals are important to analyze the physical
properties of BBG. So far the values of these atomic interactions have not
been well defined. Figure 4a is the optical-absorption spectrum calculated
with the parameters used in our work.$^{8}$ Figures 4b, 4c, and 4d are the
results calculated with three other sets of parameters$^{10,16,17}$ used in
the investigations of BBG, respectively. The optical-absorption spectra
corresponding to the four sets of parameters show similar qualitative
properties, \textit{i.e.}, they exhibit four categories of absorption peaks,
twin-peak structures, and same optical selection rules. However, the
absorption frequencies in the four results are slightly different since the
values of the four sets of parameters are distinct. The reliability of these
parameters could be verified through more accurate experimental measurements
with a purer bilayer graphene.

TB and EM methods are based on different concepts in the BBG calculations.
The former simultaneously includes a magnetic field and four important
interlayer interactions in the calculations. However, the latter takes the
LL structures of MG to study the physical properties of BBG, where
interlayer interactions are regarded as perturbations. The wave functions in
TB calculations are well characterized and different from those in EM
calculations. The magneto-optical absorption spectra studied by TB and EM
exhibit five crucial differences. The former demonstrate four categories of
absorption peaks, twin-peak structures in each category, complex optical
selection rules, and complex field-dependent absorption frequencies.
However, the latter display three peak categories, no twin-peak structure, a
single selection rule, and the simple field-dependent absorption frequencies
$\omega _{a}\propto B_{0}$ and $\omega _{a}\propto \sqrt{B_{0}}$. Moreover,
the regular frequency-dependent peak intensities are shown in TB results,
while they are not obtained in EM results. These differences mainly
originate from the different concepts and distinct definitions of the wave
functions in the two methods.

In addition to BBG, AABG also presents special optical properties, as shown
in Figures S6 and S7 of the supporting material. AABG and BBG show similar
LL features, \textit{i.e.}, two groups of LLs and an asymmetric structure.
However, these two bilayer graphenes display different optical properties.
AABG contains two categories of absorption peaks resulting from the
intragroup excitations and these peaks exhibit only one single optical
selection rule. No twin-peak structure similar to that in the BBG case
arises from the special relations in the two LL groups of AABG. The
differences of the optical absorption spectra of AABG and BBG imply that the
optical properties can not be directly identified through only the
eigenvalues of LLs. An explicit analysis of the characteristics of the wave
functions and detailed calculations of the velocity matrix elements are
necessary to calculate the optical properties. Both the magneto-electronic
and magneto-optical properties of AABG and BBG reflect the main features of
zero-field electronic and optical properties. A detailed analysis of the
magneto-optical absorption spectra in AABG is discussed in an unpublished
work (Y. H. Ho \textit{et al.}, Characterization of Landau-Level
Wavefunctions and Magneto-Optical Excitations in AA-stacked Bilayer
Graphene).

TB is widely applied to tackle physical problems of semiconductors and
carbon-related systems. Through the strategy developed by our group, the
complex Hamiltonian matrix can be solved. The eigenvalues and wave functions
are clearly depicted in this calculation, and thus can be used to interpret
the results in optical properties. Since TB calculations simultaneously
include the magnetic field and the important interlayer interactions, such
results are more reliable and accurate over a much wider frequency range.
Besides, TB can conceivably be expanded to investigate other
layered-graphene systems (multilayer graphenes and bulk graphite) and other
physical properties (electronic excitations and transport properties).

\begin{flushleft}
\textbf{SUMMARY AND CONCLUSIONS}
\end{flushleft}

In summary, it can be said that the interlayer interactions play very
important roles in the magneto-optical properties of BBG. In this work,
detailed calculations of the velocity matrix and reliable characterization
of the wave functions are provided for analyzing the optical properties.
Based on the characteristics of the wave functions, the asymmetric LLs can
be divided into two groups. These two groups lead to four categories of
prominent optical-absorption peaks $\omega _{11}$'s, $\omega _{22}$'s, $%
\omega _{12}$'s, and $\omega _{21}$'s. Each category exhibits twin-peak
structures originating from the asymmetry of LLs. The optical selection
rules of the four categories are $\triangle n_{11}=\pm 1$, $\triangle
n_{22}=\pm 1$, $\triangle n_{12}=0,-2$, and $\triangle n_{21}=0,2$
respectively. All of the selection rules can be reasonably explained by the
characteristics of the wave functions. The peak intensities of $\omega _{11}$%
's and $\omega _{22}$'s ($\omega _{12}$'s and $\omega _{21}$'s) increase
(decrease) with rising frequency. The field-dependent absorption frequencies
are not directly proportional to either $B_{0}$ or $\sqrt{B_{0}}$.
Furthermore, the convergent absorption frequencies at the weak field
strength might be helpful and reliable in determining the interlayer atomic
interactions $\gamma _{1}$ and $\gamma _{6}$. The main results of BBG in
this work are more complex in nature than those of MG, and different from
those in previously published works studied by EM. The above-mentioned
magneto-optical properties could be confirmed by magneto-absorption
spectroscopy measurements.$^{26,31,32}$

\begin{flushleft}
\textbf{METHODS}
\end{flushleft}

The optical-absorption spectrum is calculated by gradient approximation
based on the tight-binding model. The geometric configuration of a primitive
unit cell in BBG is shown in Figure 5a. The unit cell includes four
sublattices, $A_{1}$, $B_{1}$, $A_{2}$ and $B_{2}$. The subscripts 1 and 2
are, respectively, the indices of the first and second layer. The
nearest-neighbor hopping integral on the same layer is indicated by $\gamma
_{0}$ (=2.598 eV). The four important interlayer interactions$^{8}$ used in
this work are $\gamma _{1}$ (=0.364 eV), $\gamma _{3}$ (=0.319 eV), $\gamma
_{4}$ (=0.177 eV), and $\gamma _{6}$ (=-0.026 eV), indicated in Figure 5a. $%
\gamma _{6}$ is the difference of site energy between the $A$ and $B$ atoms
owing to the interlayer coupling. The perpendicular-uniform magnetic field
induces a periodic Peierls phase related to the vector potential $\mathbf{A}%
=(0,B_{0}x,0)$. The primitive unit cell shown in Figure 5b is enlarged under
this periodic condition and owns four effective sublattices including $%
2R_{B} $ $A_{1}$, $2R_{B}$ $B_{1}$, $2R_{B}$ $A_{2}$, and $2R_{B}$ $B_{1}$
atoms, respectively. $R_{B}$ is related to the dimension of the Hamiltonian
under a magnetic field;$^{12}$ for example, $R_{B}$ is 7900 for $B_{0}=10$
T. That is to say, each LL is the linear combination of the four magnetic TB
functions associated with the four effective sublattices. The representation
$A_{i,N}$ ($B_{i,N}$) indicates the $N$th ($N=1,2...2R_{B}$) $A$ ($B$) atom
on the $i$th ($i=1,2$) layer. The Hamiltonian matrix elements in the
presence of a magnetic field are
\begin{equation}
\langle \mathbf{R}_{i,N}|H_{\mathbf{B}}|\mathbf{R}_{i^{\prime },N^{\prime
}}\rangle =\gamma _{s}(\mathbf{R}_{i,N}\text{, }\mathbf{R}_{i^{\prime
},N^{\prime }})\sum \frac{1}{\text{N}}\exp [i\mathbf{k}\cdot (\mathbf{R}%
_{i^{\prime },N^{\prime }}-\mathbf{R}_{i,N})+i\frac{e}{\hbar }\Delta G(%
\mathbf{R}_{i,N}\text{, }\mathbf{R}_{i^{\prime },N^{\prime }})]\text{,}
\end{equation}%
where $\mathbf{R}_{i,N}$ is the position vector of $A_{i,N}$ or $B_{i,N}$. $%
\gamma _{s}$'s$(\mathbf{R}_{i,N}$, $\mathbf{R}_{i^{\prime },N^{\prime }})$
indicate the atomic interactions between the atoms at $\mathbf{R}_{i,N}$ and
$\mathbf{R}_{i^{\prime },N^{\prime }}$, \textit{i.e.}, they are $\gamma _{0}$%
, $\gamma _{1}$, $\gamma _{3}$, $\gamma _{4}$, and $\gamma _{6}$. $\Delta G(%
\mathbf{R}_{i,N}$, $\mathbf{R}_{i^{\prime },N^{\prime }})=\int_{0}^{1}(%
\mathbf{R}_{i^{\prime },N^{\prime }}-\mathbf{R}_{i,N})\cdot \mathbf{A}[%
\mathbf{R}_{i^{\prime },N^{\prime }}+\lambda (\mathbf{R}_{i^{\prime
},N^{\prime }}-\mathbf{R}_{i,N})]d\lambda $ is the Peierls phase induced by
a magnetic field (Please see more detailed definitions of the Hamiltonian
matrix elements in Ref. 12). After diagonalizing the Hamiltonian matrix, the
eigenvalues are obtained and thus the optical-absorption function in eq 3
can be solved through gradient approximation.

\vskip 0.6 truecm

\noindent \textit{Acknowledgments.} This work was supported by the NSC and
NCTS of Taiwan, under the grant Nos. NSC 95-2112-M-006-028-MY3 and NSC
97-2112-M-110-001-MY2.

\vskip 0.6 truecm

\noindent \textit{Supporting Information Available:} Influences of the
interlayer interactions on the magneto-electronic and magneto-optical
properties (Figures S1 and S2), Landau level and magneto-optical absorption
spectra calculated with four sets of parameters (Figures S3 and S4), optical
excitations obtained with four broadening parameters (Figure S5), and
electronic structures and optical-absorption spectra (Figures S6 and S7) of
AA-stacked bilayer graphene. This material is available free of charge
\textit{via} the Internet at http://pubs.acs.org.

\newpage
%TCIMACRO{\TeXButton{references}{\leftline {\Large \textbf {References}}}}%
%BeginExpansion
\leftline {\Large \textbf {References}}%
%EndExpansion

\begin{itemize}
\item[$^{1}$] Novoselov, K. S.; Geim, A. K.; Morozov, S. V.; Jiang, D.;
Zhang, Y.; Dubonos, S. V.; Firsov, A. A. Electric Field Effect in Atomically
Thin Carbon Films. \textit{Science} \textbf{2004}, 306, 666.

\item[$^{2}$] Novoselov, K. S.; Geim, A. K.; Morozov, S. V.; Jiang, D.;
Katsnelson, M. I.; Grigorieva, I. V.; Dubonos, S. V.; Firsov, A. A.
Two-Dimensional Gas of Massless Dirac Fermions in Graphene. \textit{Nature}
\textbf{2005}, 438, 197-200.

\item[$^{3}$] Berger, C.; Song, Z. M.; Li, T. B.; Li, X. B.; Ogbazghi, A.
Y.; Feng, R.; Dai, Z. T.; Marchenkov, A. N.; Conrad, E. H.; First, P. N.; de
Heer, W. A. Ultrathin Epitaxial Graphite: 2D Electron Gas Properties and a
Route toward Graphene-Based Nanoelectronics. \textit{J. Phys. Chem. B}
\textbf{2004}, 108, 19912-19916.

\item[$^{4}$] Rutter, G. M.; Crain, J. N.; Guisinger, N. P.; Li, T.; First,
P. N.; Stroscio, J. A. Scattering and Interference in Epitaxial Graphene.
\textit{Science} \textbf{2007}, 317, 219-222.

\item[$^{5}$] Wallace, P. R. The Band Theory of Graphite. \textit{Phys. Rev.}
\textbf{1947}, 71, 622-634.

\item[$^{6}$] Gr\"{u}neis, A.; Attaccalite, C.; Wirtz, L.; Shiozawa, H.;
Saito, R.; Pichler, T.; Rubio, A. Tight-Binding Description of the
Quasiparticle Dispersion of Graphite and Few-layer Graphene. \textit{Phys.
Rev. }B \textbf{2008}, 78, No. 205425.

\item[$^{7}$] Slonczewski, J. C.; Weiss, P. R. Band Structure of Graphite.
\textit{Phys. Rev.} \textbf{1958}, 109, 272-279.

\item[$^{8}$] Charlier, J.-C.; Michenaud, J.-P. First-Principles Study of
the Electronic Properties of Simple Hexagonal Graphite. \textit{Phys. Rev. B}
\textbf{1992}, 46, 4531-4539.

\item[$^{9}$] Latil, S.; Henrard, L. Charge Carriers in Few-Layer Graphene
Films. \textit{Phys. Rev. Lett}. \textbf{2006}, 97, No. 036803.

\item[$^{10}$] Partoens, B.; Peeters, F. M. From Graphene to Graphite:
Electronic Structure around the K Point. \textit{Phys. Rev. B} \textbf{2006}%
, 74, No. 075404.

\item[$^{11}$] Lu, C. L.; Chang, C. P.; Huang, Y. C.; Chen, R. B.; Lin, M.
F. Influence of an Electric Field on the Optical Properties of Few-Layer
Graphene with AB Stacking. \textit{Phys. Rev. B} \textbf{2006}, 73, No.
144427.

\item[$^{12}$] Lai,Y. H.; Ho, J. H.; Chang, C. P.; Lin, M. F.
Magnetoelectronic Properties of Bilayer Bernal Graphene. \textit{Phys. Rev. B%
} \textbf{2008}, 77, No. 085426.

\item[$^{13}$] McCann, E.; Fal'ko, V. I. Landau-Level Degeneracy and Quantum
Hall Effect in a Graphite Bilayer. \textit{Phys. Rev. Lett}. \textbf{2006},
96, No. 086805.

\item[$^{14}$] Abergel, D. S. L.; Fal'ko, V. I. Optical and Magneto-Optical
Far-Infrared Properties of Bilayer Graphene. \textit{Phys. Rev. B} \textbf{%
2007}, 75, No. 155430.

\item[$^{15}$] Charlier, J.-C.; Gonze, X.; Michenaud, J.-P. First-Principles
Study of the Electronic Properties of Graphite. \textit{Phys. Rev. B}
\textbf{1991}, 43, 4579-4589.

\item[$^{16}$] Nakao, J. Landau Level Structure and Magnetic Breakthrough in
Graphite. \textit{J. Phys. Soc. Japan} \textbf{1976}, 40, 761-768.

\item[$^{17}$] Kuzmenko, A. B.; Crassee, I.; van der Marel, D. Determination
of the Gate-Tunable Band Gap and Tight-Binding Parameters in Bilayer
Graphene Using Infrared Spectroscopy. \textit{Phys. Rev. B} \textbf{2009},
80, No. 165406.

\item[$^{18}$] Ohta, T.; Bostwick, A.; Seller, T.; Horn, K.; Rotenberg, E.
Controlling the Electronic Structure of Bilayer Graphene. \textit{Science}
\textbf{2006}, 313, 951-954.

\item[$^{19}$] Ohta, T.; Bostwick, A.; McChesney, J. L.; Seyller, T.; Horn,
K.; Rotenberg, E. Interlayer Interaction and Electronic Screening in
Multilayer Graphene Investigated with Angle-Resolved Photoemission
Spectroscopy. \textit{Phys. Rev. Lett}. \textbf{2007}, 98, No. 206802.

\item[$^{20}$] Novoselov, K. S.; McCann, E.; Morozov, S. V.; Fal'ko, V. I.;
Katsnelson, M. I.; Zeitler, U.; Jiang, D.; Schedin, F.; Geim, A. K.
Unconventional Quantum Hall Effect and Berry's Phase of 2$\pi $ in Bilayer
Graphene. \textit{Nat. Phys.} \textbf{2006}, 2, 177-180.

\item[$^{21}$] Zhang, Y.; Tan, Y.-W.; Stormer, H. L.; Kim, P. Experimental
Observation of the Quantum Hall Effect and Berry's Phase in Graphene.
\textit{Nature} \textbf{2005}, 438, 201-204.

\item[$^{22}$] Miller, D. L.; Kubista, K. D.; Rutter, G. M.; Ruan, M.; de
Heer, W. A.; First, P. N.; Stroscio, J. A. Observing the Quantization of
Zero Mass Carriers in Graphene. \textit{Science} \textbf{2009}, 324, 924-927.

\item[$^{23}$] Li, G.; Andrei, E. Y. Observation of Landau Levels of Dirac
Fermions in Graphite. \textit{Nat. Phys}. \textbf{2007}, 3, 623-627.

\item[$^{24}$] Matsui, T.; Kambara, H.; Niimi, Y.; Tagami, K.; Tsukada, M.;
Fukuyama, H. STS Observations of Landau Levels at Graphite Surfaces. \textit{%
Phys. Rev. Lett}. \textbf{2005}, 94, No. 226403.

\item[$^{25}$] Deacon, R. S.; Chuang, K.-C.; Nicholas, R. J.; Novoselov, K.
S.; Geim, A. K. Cyclotron Resonance Study of the Electron and Hole Velocity
in Graphene Monolayers. \textit{Phys. Rev. B} \textbf{2007}, 76, No. 081406.

\item[$^{26}$] Henriksen, E. A.; Jiang, Z.; Tung, L.-C.; Schwartz, M. E.;
Takita, M.; Wang, Y.-J.; Kim, P.; Stormer, H. L. Cyclotron Resonance in
Bilayer Graphene. \textit{Phys. Rev. Lett.} \textbf{2008}, 100, No. 087403.

\item[$^{26}$] McClure, J. W. Diamagnetism of Graphite. \textit{Phys. Rev.}
\textbf{1956}, 104, 666-671.

\item[$^{28}$] Zheng, Y.; Ando, T. Hall Conductivity of a Two-Dimensional
Graphite System. \textit{Phys. Rev. B} \textbf{2002}, 65, No. 245420.

\item[$^{29}$] Ho, J. H.; Lai, Y. H.; Chiu, Y. H.; Lin, M. F. Landau Levels
in Graphene. \textit{Physica} E \textbf{2008}, 40, 1722-1725.

\item[$^{30}$] Koshino, M.; Ando, T. Magneto-Optical Properties of
Multilayer Graphene. \textit{Phys. Rev. B} \textbf{2008}, 77, No. 115313.

\item[$^{31}$] Sadowski, M. L.; Martinez, G.; Potemski, M.; Berger, C.; de
Heer, W. A. Landau Level Spectroscopy of Ultrathin Graphite Layers. \textit{%
Phys. Rev. Lett}. \textbf{2006}, 97, No. 266405.

\item[$^{32}$] Plochocka, P.; Faugeras, C.; Orlita, M.; Sadowski, M. L.;
Martinez, G.; Potemski, M.; Goerbig, M. O.; Fuchs, J.-N.; Berger, C.; de
Heer, W. A. High-Energy Limit of Massless Dirac Fermions in Multilayer
Graphene Using Magneto-Optical Transmission Spectroscopy. \textit{Phys. Rev.
Lett}. \textbf{2008}, 100, No. 087401.

\item[$^{33}$] Benfatto, L.; Sharapov, S. G.; Carbotte, J. P. Robustness of
the Optical Conductivity Sum Rule in Bilayer Graphene. \textit{Phys. Rev.} B
\textbf{2008}, 77, No. 125422.

\item[$^{34}$] Nicol, E. J.; Carbotte, J. P. Optical Conductivity of Bilayer
Graphene With and Without an Asymmetry Gap. \textit{Phys. Rev. }B \textbf{%
2008}, 77, No. 155409.

\item[$^{35}$] Mucha-Kruczy\'{n}ski, M.; Abergel, D. S. L.; McCann, E.;
Fal'ko, V. I. On Spectral Properties of Bilayer Graphene: The Effect of an
SiC Substrate and Infrared Magneto-Spectroscopy. \textit{J. Phys.: Condens.
Matter} \textbf{2009}, 21, No. 344206.

\item[$^{36}$] Mucha-Kruczy\'{n}ski, M.; McCann, E.; Fal'ko, V. I. The
Influence of Interlayer Asymmetry on The Magnetospectroscopy of Bilayer
Graphene. \textit{Solid State Communications} \textbf{2009}, 149, 1111-1116.

\item[$^{37}$] Chang, C. P.; Lu, C. L.; Shyu, F. L.; Chen, R. B.; Fang, Y.
K.; Lin,M. F. Magnetoelectronic properties of a graphite sheet. \textit{%
Carbon} \textbf{2004}, 42, 2975-2980.

\item[$^{38}$] Chiu, Y. H.; Lai, Y. H.; Ho, J. H.; Chuu, D. S.; Lin, M. F.
Electronic Structure of a Two-Dimensional Graphene Monolayer in a Spatially
Modulated Magnetic Field: Peierls Tight-Binding Model. \textit{Phys. Rev. B}
\textbf{2008}, 77, No. 045407.

\item[$^{39}$] Ho, J. H.; Lai, Y. H.; Chiu, Y. H.; Lin, M. F. Modulation
Effects on Landau Levels in a Monolayer Graphene. \textit{Nanotechnology}
\textbf{2008}, 19, No. 035712.

\item[$^{40}$] Huang, Y. C.; Chang, C. P.; Lin, M. F. Magnetic and Quantum
Confinement Effects on Electronic and Optical Properties of Graphene
Ribbons. \textit{Nanotechnology} \textbf{2007}, 18, No. 495401.

\item[$^{41}$] Chiu, Y. H.; Ho, J. H.; Chang, C. P.; Chuu, D. S.; Lin, M. F.
Low-Frequency Magneto-Optical Excitations of a Graphene Monolayer: Peierls
Tight-Binding Model and Gradient Approximation Calculation. \textit{Phys.
Rev. B} \textbf{2008}, 78, No. 245411.

\item[$^{42}$] Huang, Y. C.; Lin, M. F.; Chang, C. P. Landau Levels and
Magneto-Optical Properties of Graphene Ribbons. \textit{J. Appl. Phys.}
\textbf{2008}, 103, No. 073709.

\item[$^{43}$] Lin, M. F.; Shung, K. W.-K. Plasmons and Optical Properties
of Carbon Nanotubes. \textit{Phys. Rev. B} \textbf{1994}, 50, 17744-17747.

\item[$^{44}$] Johnson, J. G.; Dresselhaus, G. Optical Properties of
Graphite. \textit{Phys. Rev. B} \textbf{1979}, 7, 2275-2285.

\item[$^{45}$] Chuang, K.-C.; Baker, A. M. R.; Nicholas, R. J.
Magnetoabsorption Study of Landau Levels in Graphite. \textit{Phys. Rev. B}
\textbf{2009}, 80, No. 161410(R).

%\item[$^{46}$] Ho, Y. H.; Chiu, Y. H.; Lin, M. F. Characterization of
%Landau-Level Wavefunctions and Magneto-Optical Excitations in AA-stacked
%Bilayer Graphene. (\textit{unpublished}).
\end{itemize}

\newpage \centerline {\Large \textbf {FIGURE CAPTIONS}}

\vskip0.5 truecm

Figure 1. (a) Landau levels of bilayer Bernal graphene at $B_{0}$=40 T. The
wave functions of (b) $A_{1,o}$, (c) $B_{1,o}$, (d) $A_{2,o}$, (e) $B_{2,o}$
atoms with odd integer indices are shown.

\vskip0.5 truecm

Figure 2. Optical absorption spectra of bilayer Bernal graphene at (a) 40 T,
(b) 30 T, (c) 20 T, and (d) 10 T. The spectrum of monolayer graphene at 40 T
is plotted in (e).

\vskip0.5 truecm

Figure 3. Field-dependent optical absorption frequencies of (a) bilayer
Bernal graphene and (b) of monolayer graphene.

\vskip0.5 truecm

Figure 4. (a) The optical-absorption spectra at $B_{0}=40$ T calculated with
the parameters used in our work. (b), (c), and (d) are the results
calculated with three other sets of parameters commonly used in the
investigations of bilayer Bernal graphene.

\vskip0.3 truecm

Figure 5. The primitive unit cells of bilayer Bernal graphene in the absence
and presence of a magnetic field are shown in (a) and (b) respectively. $%
\gamma _{0}$ (=2.598 eV) is the nearest-neighbor hopping integral and the
four important interlayer interactions are $\gamma _{1}$ (=0.364 eV), $%
\gamma _{3}$ (=0.319 eV), $\gamma _{4}$ (=0.177 eV), and $\gamma _{6}$
(=-0.026 eV).

\newpage
\begin{figure}
\rotatebox{0}{\includegraphics[width=1\textwidth]{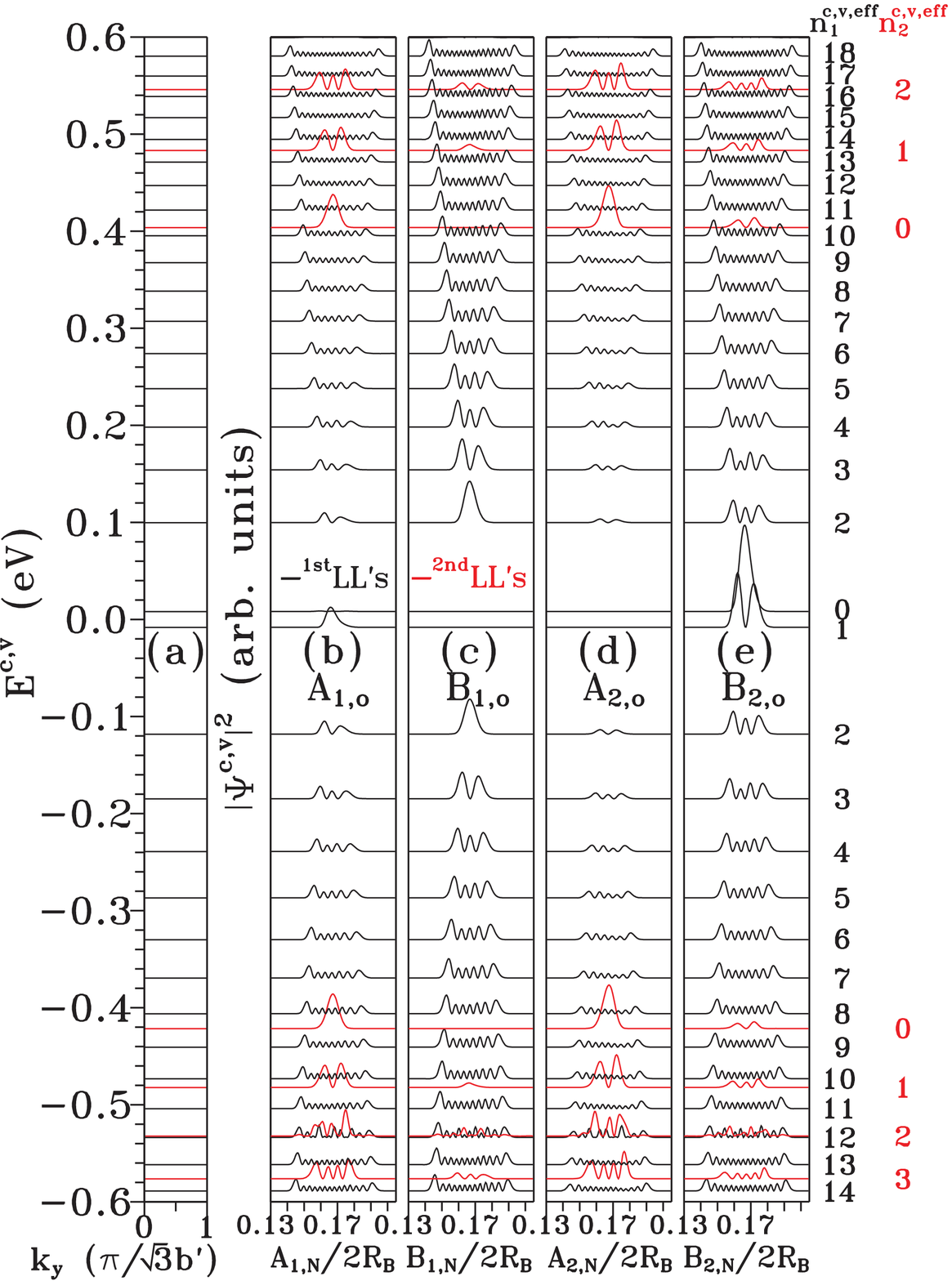}}
\end{figure}

\newpage
\begin{figure}
\rotatebox{0}{\includegraphics[width=1\textwidth]{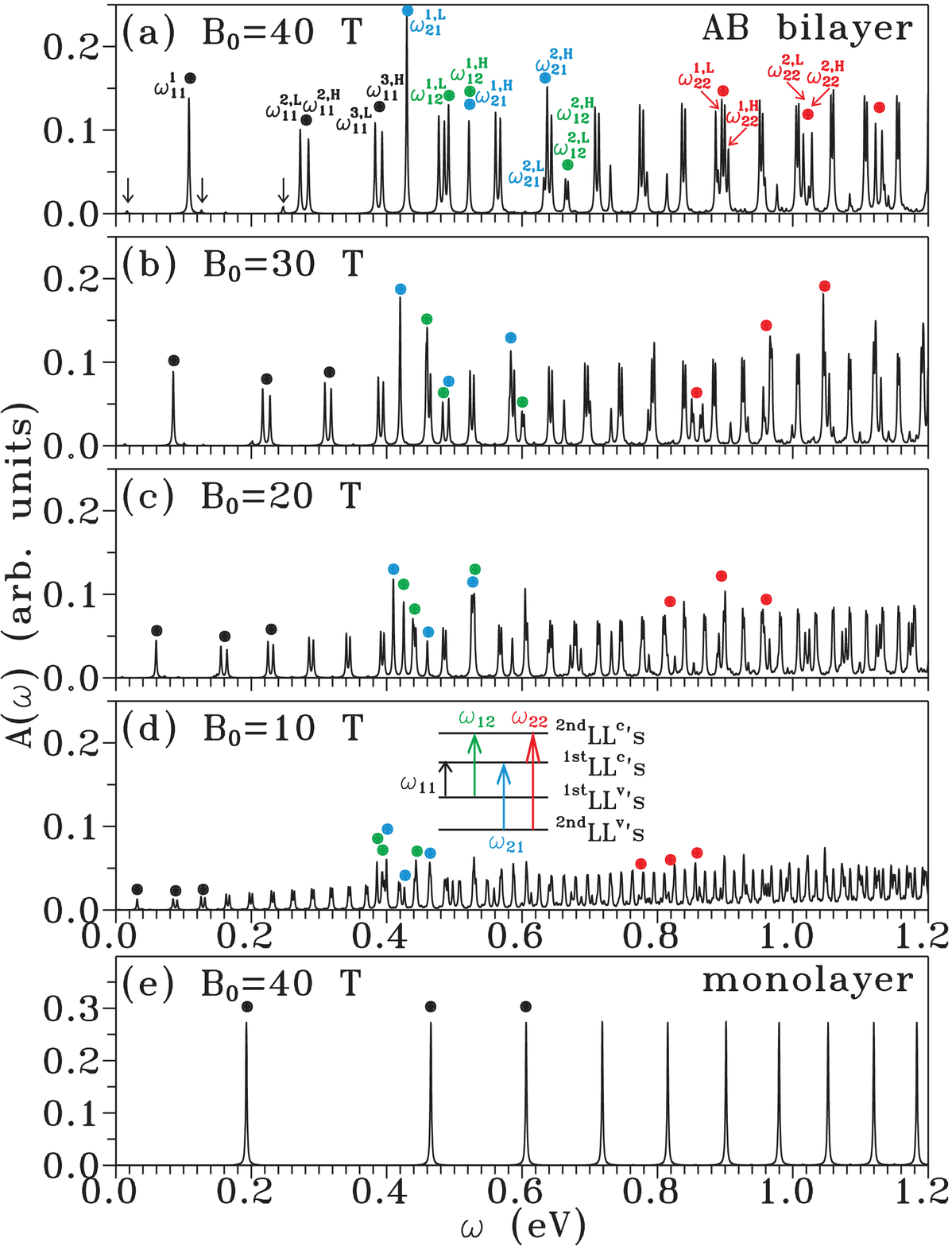}}
\end{figure}

\newpage
\begin{figure}
\rotatebox{0}{\includegraphics[width=1\textwidth]{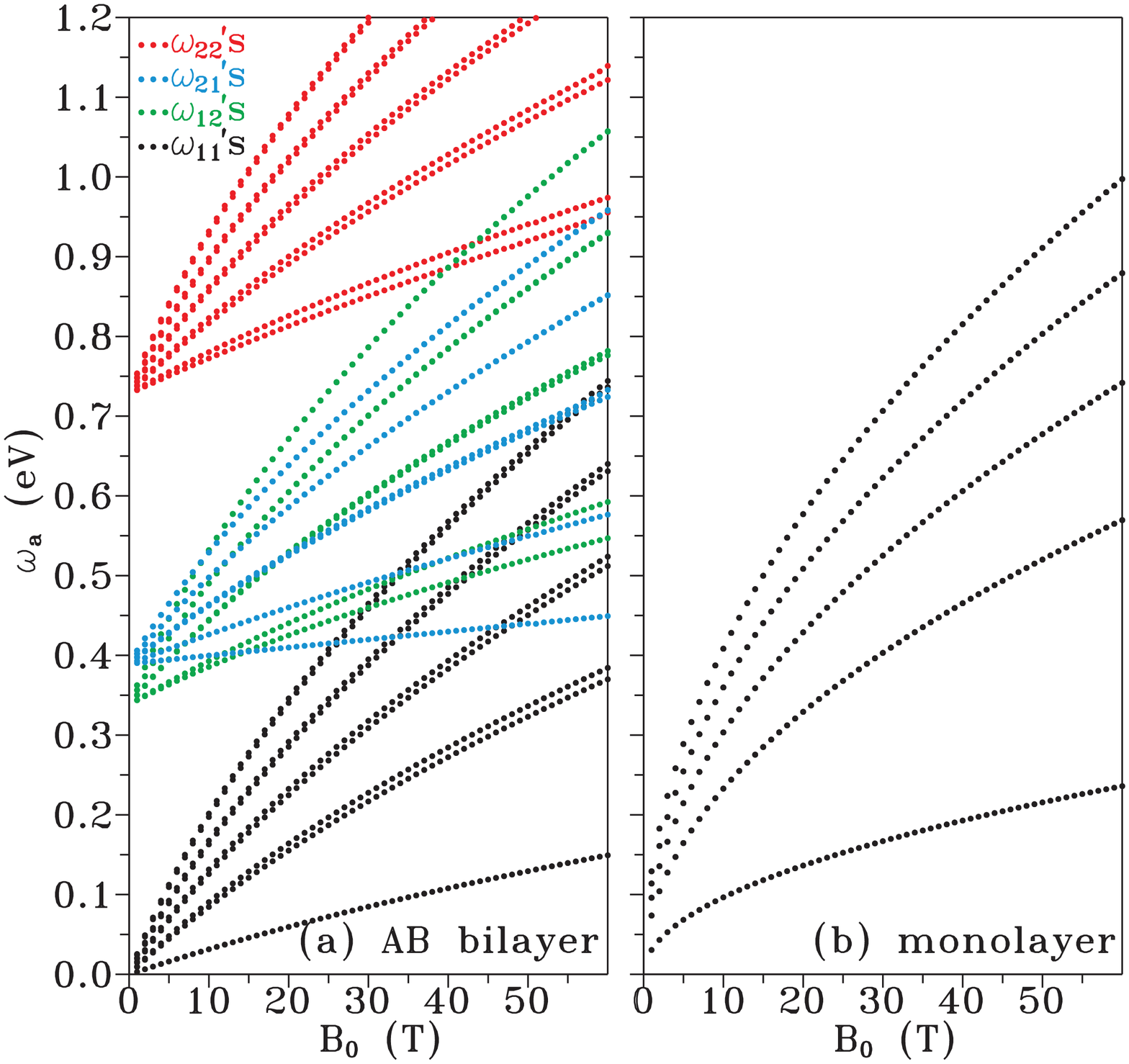}}
\end{figure}

\newpage
\begin{figure}
\rotatebox{0}{\includegraphics[width=1\textwidth]{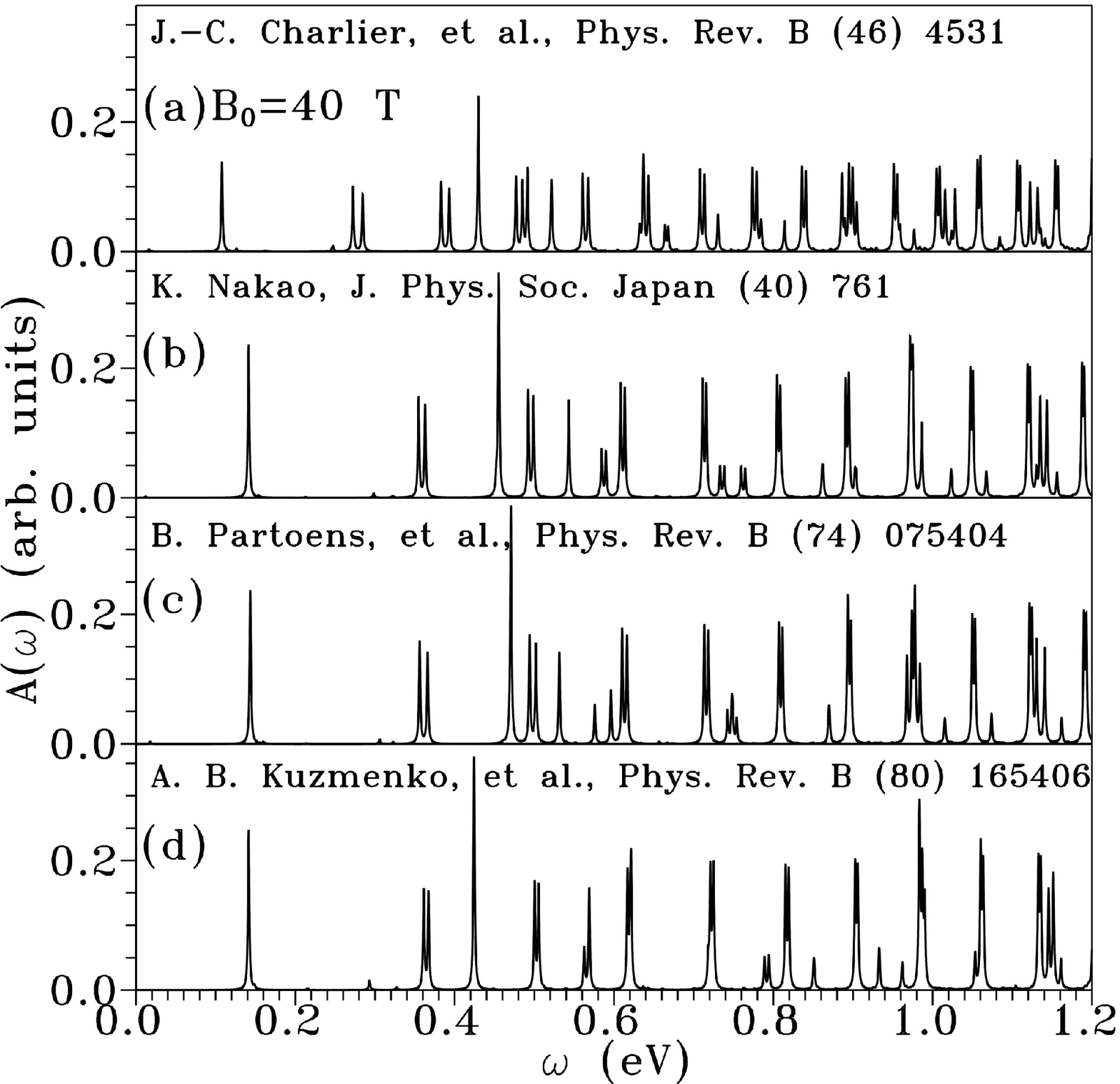}}
\end{figure}

\newpage
\begin{figure}
\rotatebox{0}{\includegraphics[width=1\textwidth]{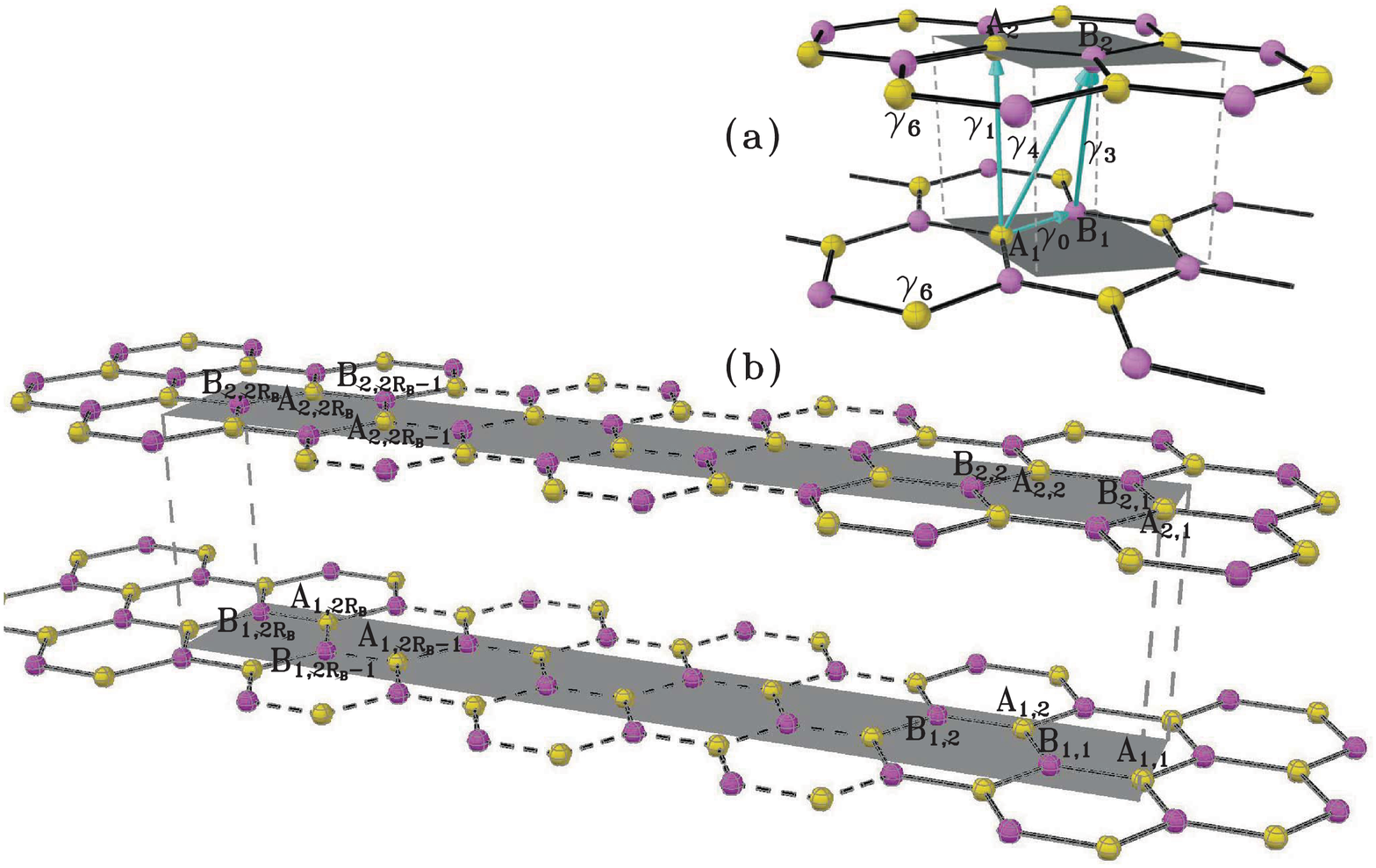}}
\end{figure}

\end{document}